\documentclass [11pt]{article}

\usepackage{amsmath,amssymb}
\usepackage{algorithm}
\usepackage{algpseudocodex}
\usepackage{graphics}
\usepackage{graphicx}
\usepackage{esvect}
\usepackage{hyperref}
\usepackage{color-edits}

\newcommand{\cev}[1]{\reflectbox{\ensuremath{\vv{\reflectbox{\ensuremath{#1}}}}}}

\addauthor[Bill]{b}{red}
\addauthor[Keld_Stefan]{k}{blue}

\begin{document}

\title{Local elimination in the traveling salesman problem}

\author{ William Cook\thanks{Supported by a Humboldt Research Award held at the Research Institute for Discrete Mathematics, University of Bonn.}\\
         Combinatorics and Optimization\\
         University of Waterloo\\
         \and
         Keld Helsgaun\\
         Department of People and Technology\\
         Roskilde University
         \and
         Stefan Hougardy\\
         Research Institute for Discrete Mathematics\\
         University of Bonn
         \and
         Rasmus T. Schroeder\\
         DB Systel GmbH, Frankfurt am Main
         }

\maketitle

\begin{abstract}
\noindent
Hougardy and Schroeder (WG 2014) proposed a combinatorial technique for pruning the search space in the traveling salesman problem, establishing that, for a given instance, certain edges cannot be present in any optimal tour.
We describe an implementation of their technique, employing an exact TSP solver to locate $k$-opt moves in the elimination process.
In our computational study, we combine LP reduced-cost elimination together with the new combinatorial algorithm.
We report results on a set of geometric instances, with the number of points $n$ ranging from 3,038 up to 115,475.
The test set includes all TSPLIB instances having at least 3,000 points, together with 250 randomly generated instances, each with 10,000 points, and three currently unsolved instances having 100,000 or more points.
In all but two of the test instances, the complete-graph edge sets were reduced to under 3$n$ edges.
For the three large unsolved instances, repeated runs of the elimination process reduced the graphs to under 2.5$n$ edges.
\end{abstract}

\section*{}

The classic Dantzig-Fulkerson-Johnson paper \cite{dfj1954} on the traveling salesman problem (TSP) introduces the cutting-plane method as an effective means for obtaining linear-programming (LP) relaxations for problems in discrete optimization and integer programming.
Dantzig et al.\ also describe briefly, in the case of the TSP, how the relaxation can be used to initiate a combinatorial procedure, aimed at reducing the search space for an optimal solution.
We study an algorithm based on this idea, proposed by Hougardy and Schroeder \cite{hs2014} in 2014.

In our work, we consider the symmetric form of the TSP, where the distance to travel from point $a$ to point $b$ is the same as the distance to travel from $b$ to $a$.
An instance of the problem can be described as a complete (undirected) graph $K_n = (V, E(K_n))$ with $n$ nodes $V$, edges $E(K_n)$, and edge lengths $(d_e: e \in E(K_n))$.
The nodes of $K_n$ represent the points to visit and the edge lengths indicate the distance to travel between the ends of each edge.
The problem asks to find a {\em tour} of minimum total edge length, where a tour is a circuit of $K_n$ visiting each node exactly once, also known as a {\it Hamiltonian circuit}.

In the class of LP relaxations considered by  Dantzig et al., a tour $T$ with edge set $E(T) \subseteq E(K_n)$ is represented in variables $(x_e: e \in E(K_n))$ by setting $x_e = 1$ for each $e \in E(T$) and $x_e = 0$ for each $e \notin E(T)$; constraints are linear equations and inequalities satisfied by all such tour vectors; the objective function is to minimize $\sum(d_ex_e: e \in E(K_n))$.
A dual solution to such an LP relaxation provides a lower bound on the length of any tour for the given TSP instance. 
Letting $L$ denote the value of the lower bound and $U$ the length of a known TSP tour, then $\Delta \equiv U - L \geq 0$ is called the {\em integrality gap} with respect to the relaxation and tour.
An important observation of Dantzig et al.\ \cite{dfj1954} is that if a variable $x_e$ has reduced cost in the LP dual solution greater than $\Delta$, then $x_e$ can be eliminated from the problem, since the corresponding edge $e$ cannot be included in any optimal TSP tour.
Similarly,  if the reduced cost of $x_e$ is less than $-\Delta$, then the value of $x_e$ can be fixed to 1, since edge $e$ must be present in every optimal tour. 
Concerning this, Dantzig et al.\ \cite{dfj1954} write the following, where $E$ is used in place of $\Delta$.
\begin{quote}
During the early stages of the computation, $E$ may be quite large and very few links can be dropped by this rule; however, in the latter stages often so many links are eliminated that one can list all possible tours that use the remaining admissible links.
\end{quote}
In 1959, they published a second paper \cite{dfj1959} on the topic, writing ``it is our belief that a linear-programming, combinatorial approach affords a practical way of solving traveling-salesman problems'' but note in conclusion that they did not ``indicate how one could make the combinatorial analysis a routine procedure.''

Reduced-cost elimination and variable fixing are now standard in TSP computation, as well as in general integer programming and discrete optimization.
In comparison, little attention has been given to further restricting the solution space via combinatorial analysis of the type proposed by Dantzig et al.
The goal of this paper is to demonstrate, within the context of the TSP, the potential power of the combination of the two techniques.

The combinatorial algorithm we study is an edge-elimination/fixing method developed by Hougardy and Schroeder \cite{hs2014}.
The central idea was proposed by Jonker and Volgenant \cite{jv1984}, in the context of a spanning-tree-based branch-and-bound algorithm for the TSP.
Jonker and Volgenant write the following.
\begin{quote}
The simple identification rules are based on the fact that a solution can be optimal only if it is 2-optimal.
\end{quote}
That is, they eliminate certain edges $e$ by proving any tour $T$ containing $e$ can be improved by exchanging two edges.
Hougardy and Schroeder extend this to general local-improvement moves, possibly exchanging large sets of edges in a non-optimality proof.

In Section~\ref{section_local_elimination}, we present Hougardy-Schroeder's method as a 2-person game, between an ``edge eliminator'' and a ``tour builder.'' 
Edge $e$ is eliminated/fixed by demonstrating a winning strategy for the eliminator.
Details of our engineering steps to implement this {\it local elimination} process are given in Section~\ref{section_implementation}.
In Section~\ref{section_computation}, we report on a computational study of geometric instances, with the number of points $n$ ranging from 3,038 up to 115,475, including all TSPLIB instances having at least 3,000 points.
In 14 of the 16 examples, the complete-graph edge sets were reduced to under 3$n$ edges, and also in 14 of the 16 examples we fixed at least 0.1$n$ edges.
In Section~\ref{section_certification}, we discuss a data structure to record the elimination process, certifying the correctness of the computations.
We present such certificates for reduced edge sets of the three open TSP instances in our test set, each having 100,000 or more points.
In each of the three cases, repeated runs of the elimination process reduced the edge sets to under 2.5$n$ edges.
The certificates allow the elimination results to potentially be employed in an exact solver, as we discuss in Section~\ref{section_remarks}.

\section{Local elimination}
\label{section_local_elimination}

Suppose we have an instance of the TSP specified by $K_n$ and edge lengths $(d_e: e \in E(K_n))$.
Rather than working with the full edge set, we consider a graph $G = (V, E)$ where $E \subseteq E(K_n)$ and $E$ contains all optimal tours.
For example, as in our computational study, $E$ could be the set of edges that remain after LP reduced-cost elimination is applied to the full instance.
For $u, v \in V$, we denote by $uv$ the edge having ends $\{u, v\}$ and we say $u$ is a {\em neighbor} of $v$ if $uv \in E$.

We will discuss a technique for proving an edge $e = ab \in E$ is not contained in any optimal TSP tour.

\subsection{Witnesses}
\label{section_witnesses}

Let ${\cal Q} = \{F_1, F_2, \ldots, F_q\}$ be a family of edge sets such that, for each $i = 1, \ldots, q$, $F_i \subseteq E$ and $e \in F_i$.
We call ${\cal Q}$ a {\em witness family} for edge $e$, or an $e$-{\em witness family}, if it has the property that every tour $T$ of $G$ having $e \in E(T)$ also has $F_i \subseteq E(T)$ for some $i \in \{1, \ldots, q\}$.

\begin{itemize}
\item Example 1: A simple $e$-witness family is to take, for a positive integer $t < n$, the edge sets of all paths of length $t$ containing edge $e$ in the ``center'' of the path.
To be precise, we say a path $P$ with ordered edges $(e_1, e_2, \ldots , e_t)$ is $e$-{\it centered} if $e_{\lceil t/2 \rceil} = e$, where $\lceil t/2 \rceil$ denotes $t/2$ rounded up to the nearest integer.
A tour $T$ with $e \in E(T)$ must contain an $e$-centered path for any specified $t$ such that $1 \leq t \leq n.$

\item Example 2: Every tour $T$ of $G$ with $e = ab \in E(T)$ must contain a second edge meeting node $a$.
Therefore, another simple $e$-witness family consists of the sets $\{e, ax\}$ for each neighbor $x$ of $a$ such that $x \neq b$.
%Therefore, another simple witness family for $e$ consists of the sets $\{e, e_i\}$ for each $e_i \in \delta_G(a) \setminus \{e\}$.

\item Example 3: Similarly, for any $v \in V\setminus \{a, b\}$, we obtain an $e$-witness family by taking $\{e, vx, vy\}$ all for pairs $x, y$ of neighbors of $v$.
\end{itemize}

\noindent
We will use these three examples as building blocks to construct complex witness families.
The goal of the construction process is to produce (possibly large) edge sets $F$ that are {\it incompatible} with TSP optimality, that is, there does not exist an optimal tour $T$ with $F \subseteq E(T)$.
If edge $e$ has a witness family ${\cal Q}$ such that each $F \in {\cal Q}$ is incompatible with TSP optimality, then $e$ can be eliminated from the TSP instance.

\subsection{Nowhere $k$-optimality}
\label{section_nowhere}

A $k$-{\it opt move}, for a tour $T$ and integer $k \geq 2$, consists of replacing $k$ edges in $T$ to obtain a shorter tour $T'$.
That is, we delete $k$ edges and reconnect the resulting paths into a tour, where the set of $k$ edges added to reconnect the paths has total length less than that of the deleted edges.
For convenience, we permit the case where a deleted edge is one of those added back, so a $k$-opt move may also be an $l$-opt move for some $l < k$.

A $k$-opt move certifies tour $T$ is not optimal.
We make repeated use of this simple fact in a certificate showing that $F \subseteq E$ is incompatible with TSP optimality.

In this discussion, it is convenient to express $F$ as the union of edge sets of node-disjoint paths $p_1, \ldots , p_m$.
We call $P = \{p_1, p_2, \ldots , p_m\}$ a {\em path system}.
For each $i = 1, \ldots , m$, let $\vv{p}_i$ denote path $p_i$ oriented from one end node $s_i$ to the other end node $t_i$, and let $\cev{p}_i$ denote the reverse path, from $t_i$ to $s_i$.

\begin{figure}[htb]
  \centering
  \includegraphics[width=5.0in]{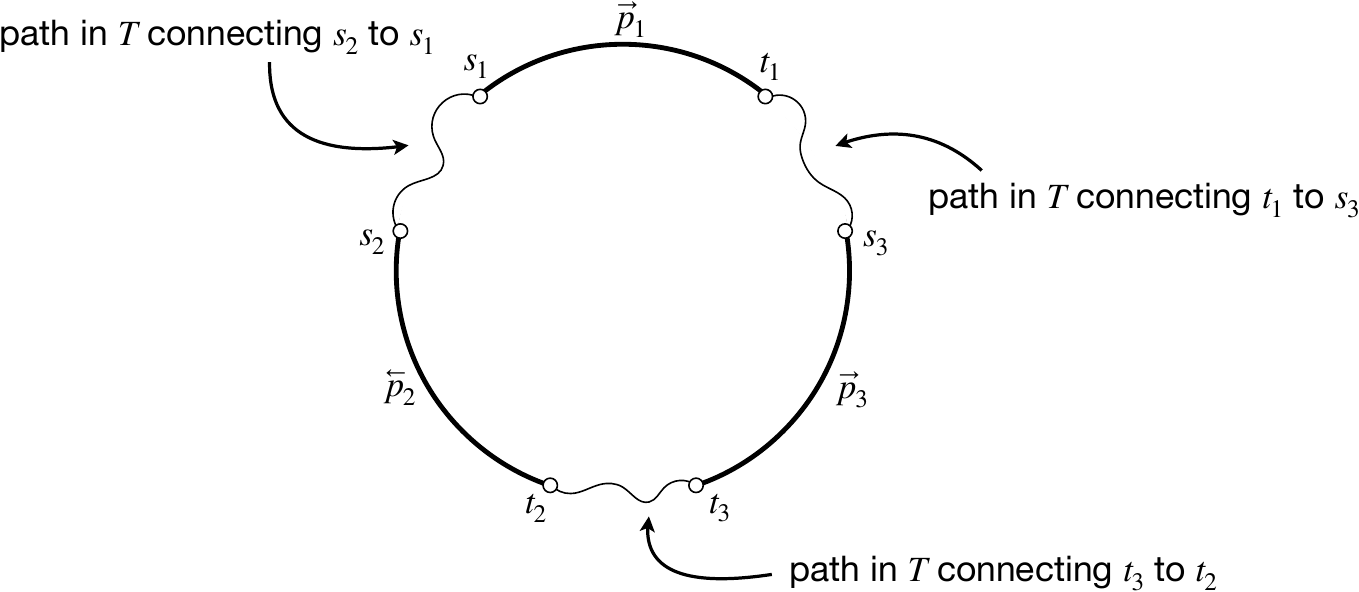}
  \caption{Tour $T$ traversing the ordered path system $[\protect\vv{p}_1, \protect\vv{p}_3, \protect\cev{p}_2]$.}
  \label{figpath1}
\end{figure}

Letting ${\cal T}_F$ denote the set of tours of $G$ containing $F$, then an orientation of any $T \in {\cal T}_F$ will visit consecutively the edges in each path $p_i \in P$, either in the  $\vv{p}_i$ or $\cev{p}_i$ direction.
To fix the orientation of $T$, we select $s_1$ as the tour's start node and orient $T$ to begin with a traversal of $\vv{p}_1$.
The tour then traverses $P$ in the {\it path ordering}
$$[\vv{p}_1, \overline{p}_{\sigma(2)}, \ldots , \overline{p}_{\sigma(m)}]$$
where $\sigma$ is a permutation of $\{1,\ldots,m\}$ with $\sigma(1) = 1$ and $\overline{p}_i$ denotes either $\vv{p}_i$ or $\cev{p}_i$, for each $i = 2,\ldots,m.$
An illustration of such a traversal is given in Figure~\ref{figpath1}.

\begin{figure}[htb]
  \centering
  \includegraphics[width=1.8in]{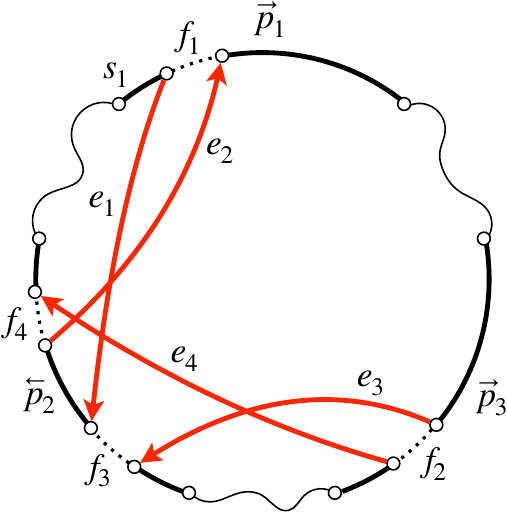}
  \caption{4-opt move, deleting $f_1,f_2, f_3, f_4 \in F$ and adding edges $e_1, e_2, e_3,e_4$.}
  \label{fig4opt}
\end{figure}

We partition ${\cal T}_F$ according to each tour's traversal of the path system $P$.
Thus there are $2^{m-1}(m-1)!$ classes in the partition, one for each possible path ordering.
Let $\cal{C}$ be a class in the partition and consider a tour $T \in {\cal C}$.
Observe that a $k$-opt move for $T$ such that all deleted edges belong to $F$ is also a $k$-opt move for all other tours in ${\cal C}$, since the validity for the $k$-opt move does not depend on the segments of $T$ that connect the end nodes of the path ordering.
See the illustration in Figure~\ref{fig4opt}.

We call $F \subseteq E$ {\em nowhere $k$-optimal} if each class of tours in the path-ordering partition has a $k$-opt move with all deleted edges in $F$.
Finding such a collection of $k$-opt moves is our mechanism for showing $F$ is incompatible with TSP optimality.
The fact that the edges in the $k$-opt moves are entirely among the nodes in $F$ allows the elimination algorithm to work locally in the graph $G$, permitting the application of the process on large-scale instances of the TSP.

\subsection{Hamilton versus Tutte}
\label{section_ht}

The complexity of proving a set $F$ is nowhere $k$-optimal grows rapidly with $|F|$ and with the number of paths in the corresponding path system $P_F$.
Thus the elimination procedure aims to build a witness family ${\cal Q}$ for a target edge $e \in E$ such that each $F \in {\cal Q}$ has small values for these two measures.

The process is akin to a game, where one player, {\em Hamilton}, reveals edges purported to be in an optimal tour and the second player, {\em Tutte}, attempts to show the revealed edge set is nowhere $k$-optimal.
As the first move, Tutte selects an integer $t \geq 1$ and requests Hamilton reveal the edges $F$ in an $e$-centered path of length $t$ containing $e$.
In each subsequent move, Tutte selects a node $v \in V$ and requests Hamilton reveal two edges meeting $v$. 
If $v$ is an end node of a path in $P_F$, then Hamilton's reveal adds a single edge to $F$.
Otherwise, Hamilton's reveal adds two edges to $F$.
The game ends in one of two ways.
Tutte wins if $F$ is shown to be nowhere $k$-optimal, while Hamilton wins if the complexity of $F$, measured as function of $|F|$ and $|P_F|$, exceeds a prescribed threshold.

To eliminate edge $e$, we show that Tutte has a winning strategy.
That is, we prepare a response to each possible Hamilton reveal.
\begin{itemize}
\item
In the first move, we consider every extension of $e$ to an $e$-centered path of length $t$ in graph $G$.
\item In a Tutte move where the selected $v$ is an end node in $P_F$, we consider every extension of the path containing $v$ by a single edge $(v, w) \in E$.
\item In a Tutte move where the selected $v$ is not part of $P_F$, we consider each pair of edges $(u,v), (v,w) \in E$.
\end{itemize}
Tying this back to our earlier discussion, note that the Hamilton reveals follow the witness building blocks described in Section~\ref{section_witnesses}.

The steps of a game strategy can be recorded in a tree, where non-leaf tree nodes are associated with Tutte moves and tree edges are associated with the possible Hamilton reveals.
Thus a non-leaf tree node has a child for every Hamilton reveal that can be made in response to the specified Tutte move.
Each node $x$ of the tree has a corresponding edge set $F_x$, obtained as the union of the Hamilton reveals associated with the edges on the path from $x$ up to the root of the tree.
In a winning strategy, each $F_x$ is below the complexity threshold and for leaf nodes $x$, the set $F_x$ is shown to be nowhere $k$-optimal.
The corresponding witness family is 
$${\cal Q} = \{F_x: x \mbox{\ is a leaf node of the game tree}\}.$$

\begin{figure}[htb]
  \centering
  \includegraphics[width=5.0in]{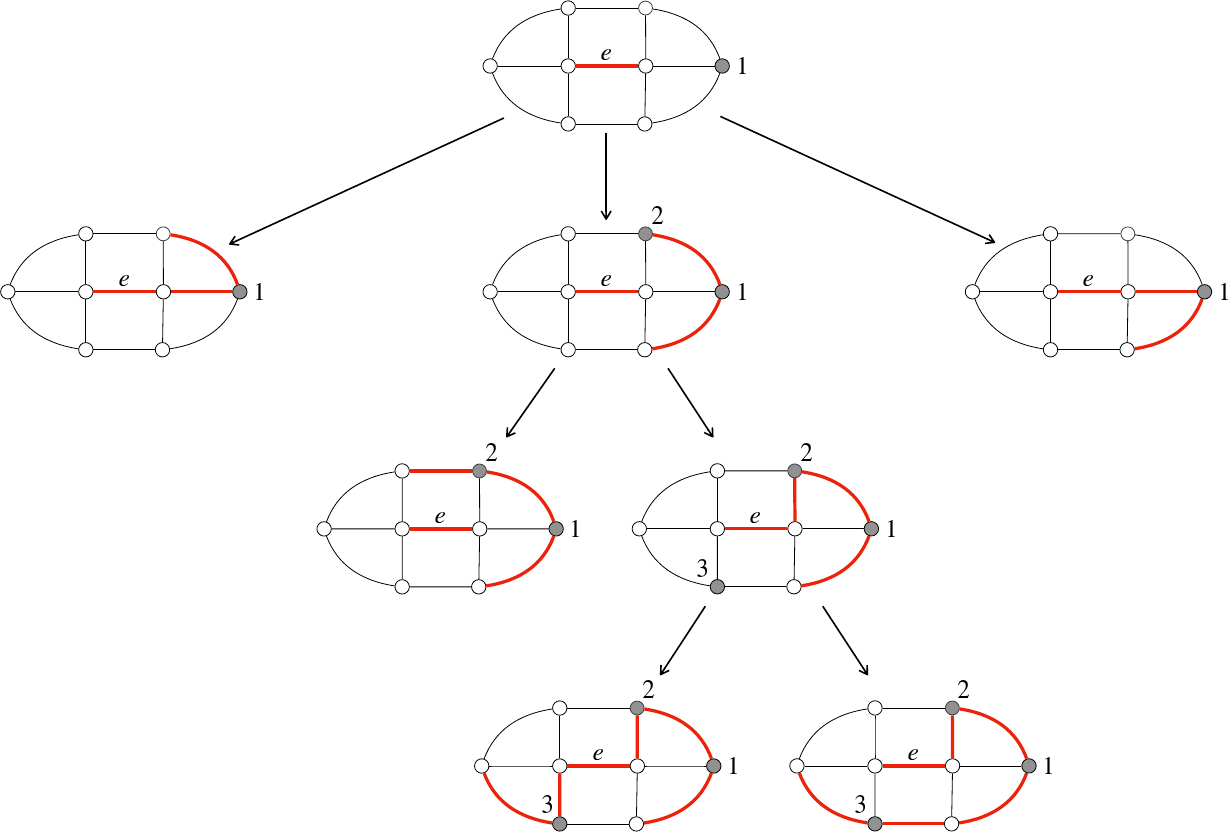}
  \caption{Hamilton-Tutte game tree with initial path length $t = 1$.}
  \label{fig_ht}
\end{figure}

An example of a Hamilton-Tutte game tree is illustrated in Figure~\ref{fig_ht}.
In this toy instance, the target edge $e$ is in the center of the graph and the initial Tutte move requests only the path of length $t = 1$.
Subsequent Tutte moves are indicated by the grey nodes, numbered 1, 2, and 3.
At each node of the tree, the union of the Hamilton reveals (the sets $F_x$) are indicated by the thick (red) edges.
There are five leaf nodes where the indicated edge sets would need to be shown to be nowhere $k$-optimal.

In general, to find a winning game tree, we adopt a backtracking search procedure, where we consider a small number of candidate Tutte moves at each internal tree node $x$.
While processing $x$, if a candidate Tutte move leads to a Hamilton reveal exceeding the complexity threshold, we consider $x$'s next Tutte candidate.
If we exhaust the supply of candidates at $x$, we backtrack to $x$'s parent node in the game tree, and consider the next Tutte candidate for the parent.

The tree-building process is outlined in pseudocode in Algorithm~\ref{algo_HT}.
The main function, \textsc{HamiltonTutte}, takes as input $e \in E$ and a length $t$ for the initial $e$-centered path to be revealed by Hamilton.
Backtracking is handled by recursive calls to a function \textsc{TutteMove}, that considers one by one the list of Tutte candidates for an edge set $F$.

\begin{algorithm}[htb]
\caption{\ Building a Hamilton-Tutte tree.}
\label{algo_HT}
\begin{algorithmic}
\State \

\Function{HamiltonTutte}{$e$, $t$} \Comment{$e$ an edge, $t$ an integer}
  \For{each $e$-centered $t$-length path $P$ containing edge $e$}
    \If{\Call{TutteMove}{$E(P)$} returns {\tt Failure}}
      \State \Return {\tt Failure}
    \EndIf
  \EndFor
  \State \Return {\tt Success}
\EndFunction
\State

\Function{TutteMove}{$F$} \Comment{$F$ an edge set}
  \If{$F$ exceeds the complexity threshold}
    \State \Return {\tt Failure}
  \EndIf
  \If{$F$ is nowhere $k$-optimal}
    \State \Return {\tt Success}
  \EndIf
  \For{each candidate Tutte move $v$}
   \If{\Call{HamiltonReveal}{$F, v$} returns {\tt Success}}
     \State{\Return {\tt Success}}
   \EndIf
  \EndFor
  \State \Return {\tt Failure}
\EndFunction
\State

\Function{HamiltonReveal}{$F, v$} \Comment{$F$ an edge set, $v$ a node}
  \For{each Hamilton reveal $H$ for node $v$}
    \If{\Call{TutteMove}{$F \cup H$} returns {\tt Failure}}
       \State \Return {\tt Failure}
    \EndIf
  \EndFor
  \State \Return {\tt Success}
\EndFunction
\State
\end{algorithmic}
\end{algorithm}

This is a rather simple strategy for producing the witness family ${\cal Q}$, but it has flexibility in its possible implementation.
For example, short searches can be used in a pre-processing step to potentially improve the practical performance of TSP algorithms for modest-sized instances, such as in the original work of Jonker and Volgenant \cite{jv1984}.
While long searches give the possibility to reduce the search space to such a degree that new techniques can be considered in the exact solution of very large instances.

\section{Implementation}
\label{section_implementation}
%%
%%
%%

%The process described in Section~\ref{section_local_elimination} is powerful in its ability to eliminate edges from the TSP variable space, but the search for a winning Hamilton-Tutte tree can lead to long computation times, making the technique impractical in potential applied settings.
%In this section, we present general ideas for improving the search time, together with specific choices made in our implementation of these ideas.

In this section, we present general ideas for improving the time to search for a winning Hamilton-Tutte tree, together with specific choices made in our implementation of these ideas.

\subsection{Verifying nowhere $k$-optimality}

We begin with techniques for showing an edge set $F \subseteq E$ is nowhere $k$-optimal, and thus incompatible with TSP optimality.

\subsubsection{Reusing $k$-opt moves}

\begin{figure}[htb]
  \centering
  \includegraphics[height=2.2in]{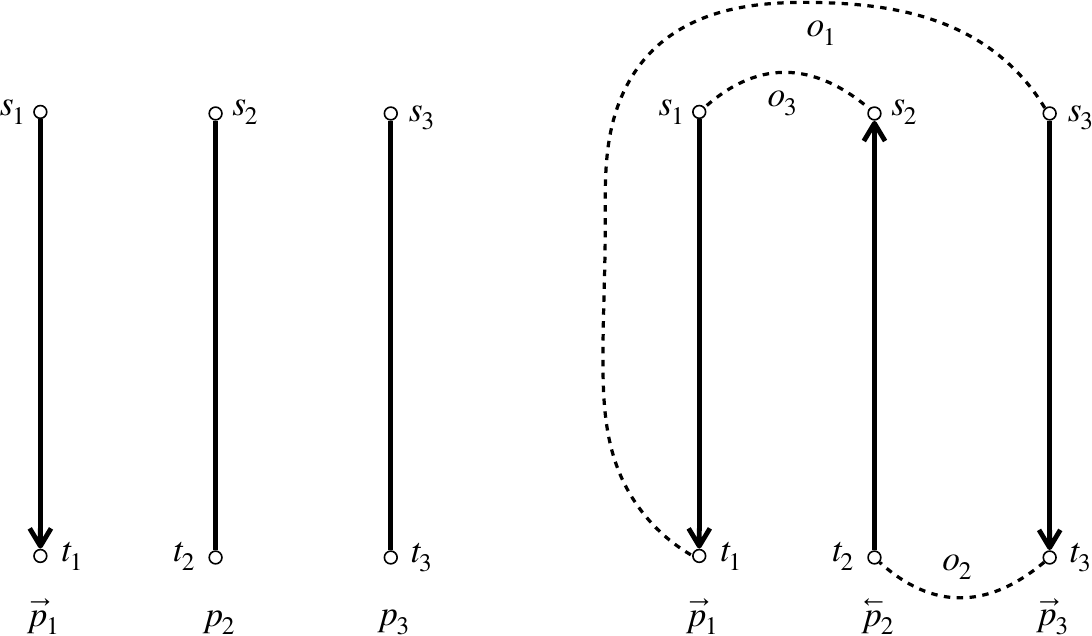}
  \caption{Outside matching $\{o_1, o_2, o_3\}$ and ordered path system $[\protect\vv{p}_1, \protect\vv{p}_3, \protect\cev{p}_2]$.}
  \label{fig_outside}
\end{figure}

As in the previous section, we express $F$ as a path system $P_F = \{p_1, p_2, \ldots , p_m\}$, where for each $i = 1,\ldots, m$ we let $s_i, t_i$ denote the end nodes of path $p_i$.
To show $F$ is nowhere $k$-optimal, we must consider all path orderings
$$P_F(\sigma, {\bf b}) \equiv [\vv{p}_1, \overline{p}_{\sigma(2)}, \ldots , \overline{p}_{\sigma(m)}]$$
where $\sigma$ is a permutation of $\{1,\ldots,m\}$ with $\sigma(1) = 1$, ${\bf b}$ is an m-dimensional binary vector with ${\bf b}[1] = 0$, and for $i = 2,\ldots,m$
\begin{equation*}
  \overline{p}_i \equiv
    \begin{cases}
       \vv{p}_i  & \text{if ${\bf b}[i] = 0$}\\
       \cev{p}_i & \text{if ${\bf b}[i] = 1.$}
    \end{cases}       
\end{equation*}
Notice that $P_F(\sigma, {\bf b})$ can be specified by listing the edges $\{o_1, \ldots , o_m\}$ used to join the paths into a circuit following the prescribed ordering.
To be precise, for $i = 1,\ldots,m$, we let
\begin{equation*}
  o_i \equiv
    \begin{cases}
       (t_{\sigma(i)}, s_{\sigma(i+1)})  & \text{if ${\bf b}[\sigma(i)] = 0$ and $ {\bf b}[\sigma(i+1)] = 0$}\\
       (t_{\sigma(i)}, t_{\sigma(i+1)})  & \text{if ${\bf b}[\sigma(i)] = 0$ and $ {\bf b}[\sigma(i+1)] = 1$}\\
       (s_{\sigma(i)}, s_{\sigma(i+1)})  & \text{if ${\bf b}[\sigma(i)] = 1$ and $ {\bf b}[\sigma(i+1)] = 0$}\\
       (s_{\sigma(i)}, t_{\sigma(i+1)})  & \text{if ${\bf b}[\sigma(i)] = 1$ and $ {\bf b}[\sigma(i+1)] = 1$}
    \end{cases}       
\end{equation*}
where $\sigma(m+1) = 1$.
We refer to this as the {\em outside matching} for $P_F(\sigma, {\bf b})$ and denote the edge set by $O_F(\sigma, {\bf b})$.
In Figure~\ref{fig_outside} we illustrate the outside matching for the tour displayed in Figure~\ref{figpath1}; in this example $\sigma = (1,3,2)$ and ${\bf{b}} = [0,1,0]$.

By construction, adding $O_F(\sigma, {\bf b})$ to $F$ gives a TSP tour through the union of the ends of $F$; we denote this tour by $T_F(\sigma, {\bf b})$.
For all choices of $\sigma$ and ${\bf b}$, we must find a $k$-opt move for $T_F(\sigma, {\bf b})$ that deletes only edges in $F$, leaving the outside matching in the tour.
We discuss below the techniques we use to find such a $k$-opt move.
Here we observe that, once found, a single $k$-opt move can possibly be applied to handle more than one choice of $\sigma$ and ${\bf b}$.

\begin{figure}[htb]
  \centering
  \includegraphics[height=2.2in]{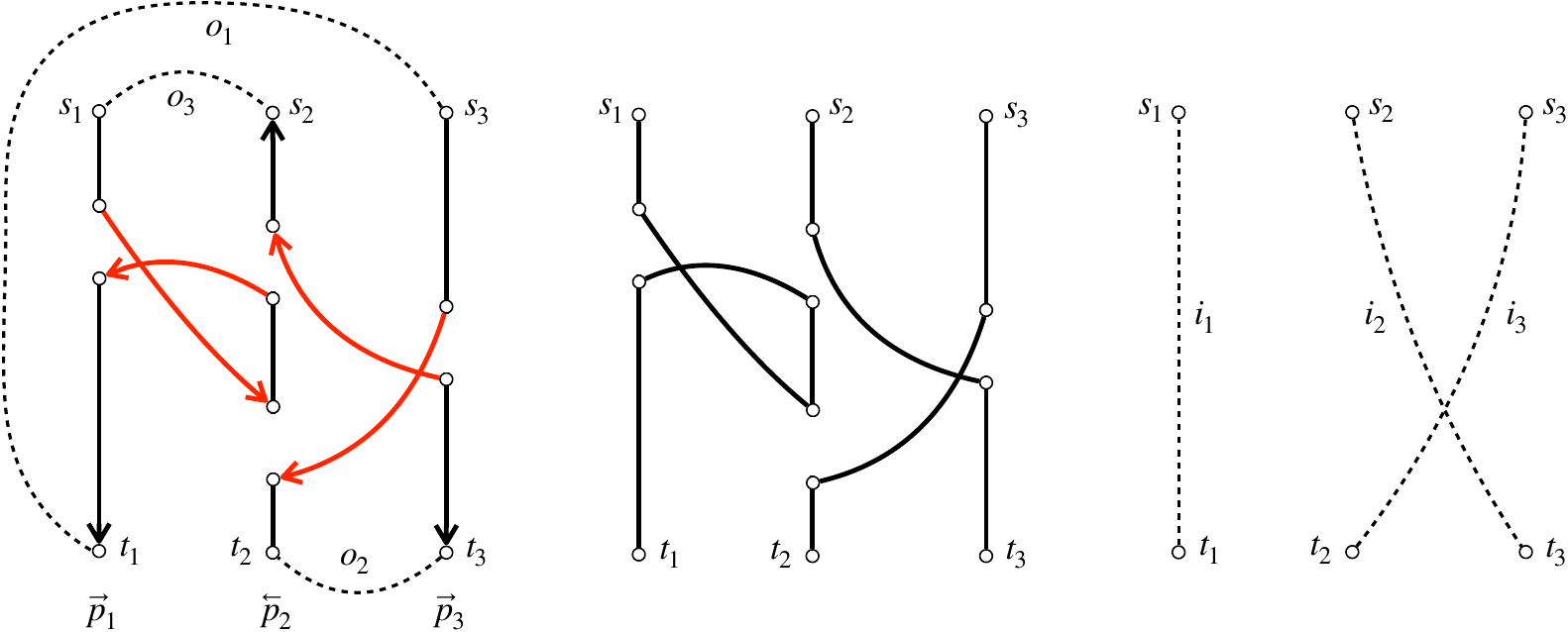}
  \caption{$4$-opt move and inside matching $\{i_1, i_2, i_3\}$.}
  \label{fig_inside}
\end{figure}

Indeed, consider a $k$-opt move for the tour $T_F(\sigma, {\bf b})$ that deletes only edges in $F$ and let $T'$ denote the shorter tour produced by the move.
Deleting the outside matching $O_F(\sigma, {\bf b})$ from $T'$ leaves node-disjoint paths $q_1, \ldots, q_m$ that together contain all nodes in $T_F(\sigma, {\bf b})$.
Each path $q_j$ starts and ends at nodes $v_{j_1}$ and $v_{j_2}$ from the set $\{s_1, t_1, \ldots, s_m, t_m\}$.
Letting $i_j $ denote the edge $v_{j_1}v_{j_2}$ for each $j = 1, \ldots, m$, we call $\{i_1, \ldots, i_m\}$ the {\em inside matching} for the $k$-opt move.
This construction is illustrated in Figure~\ref{fig_inside} for a 4-opt move applied to the example displayed in Figure~\ref{fig_outside}.

\begin{figure}[htb]
  \centering
  \includegraphics[height=2.2in]{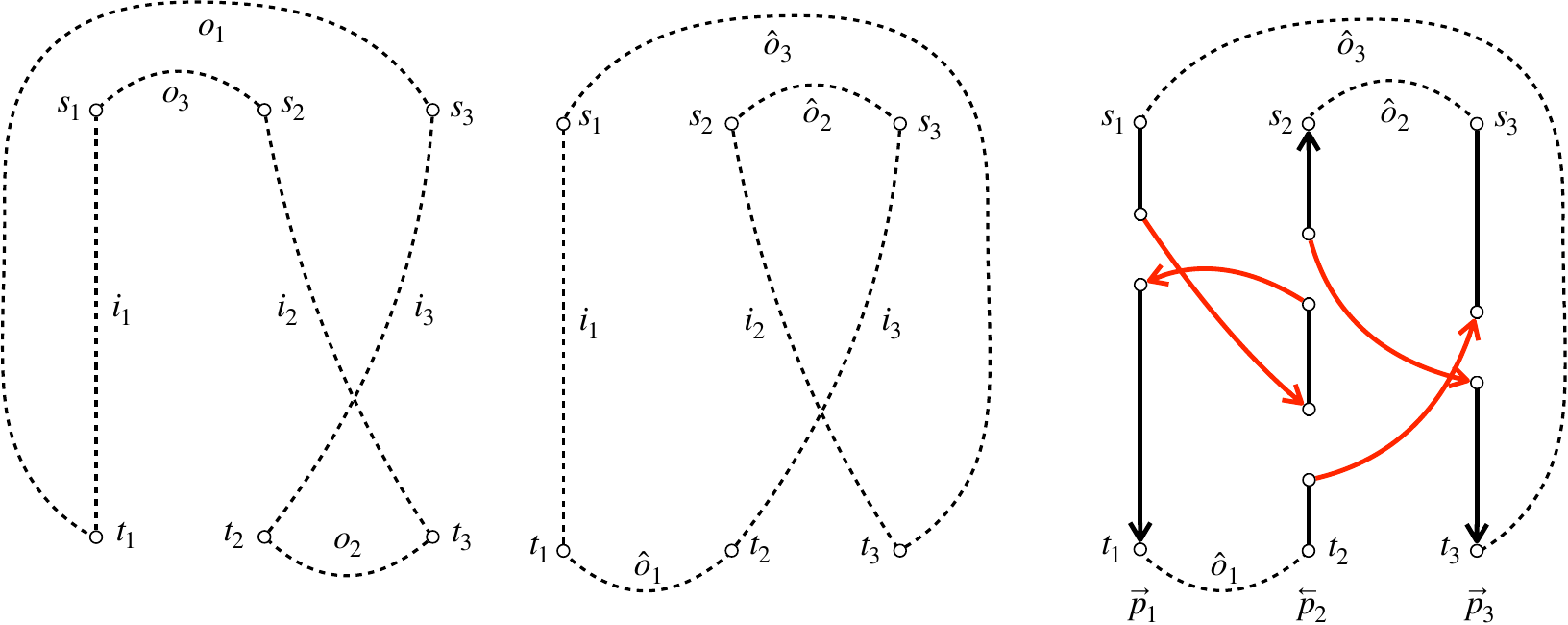}
  \caption{Outside matching $\{\hat{o}_1, \hat{o}_2, \hat{o}_3\}$ compatible with $\{i_1, i_2, i_3\}$.}
  \label{fig_outside2}
\end{figure}

The union of two matchings $\{o_1, \ldots, o_m\}$ and $\{i_1, \ldots, i_m\}$ forms a circuit through the nodes $\{s_1, t_1, \ldots, s_m, t_m\}$.
Moreover, the $k$-opt move that produced $T'$ can be applied to any tour $T_F(\hat{\sigma}, {\bf \hat{b}})$ such that the ordered path system $P_F(\hat{\sigma}, \hat{\bf b})$ corresponds to an outside matching $\{\hat{o}_1, \ldots, \hat{o}_m\}$ that forms a circuit with $\{i_1, \ldots, i_m\}$.
Such a second outside matching is illustrated in Figure~\ref{fig_outside2}, corresponding to the ordered path system $[\protect\vv{p}_1, \protect\cev{p}_2, \protect\vv{p}_3]$.

We make direct use of this observation, creating a list of the $2^{m-1}(m-1)!$ possible outside matchings to process, corresponding to all choices of $\sigma$ and ${\bf b}$.
When a $k$-opt move is found, we extract the inside matching $\{i_1, \ldots, i_m\}$ and remove from the list all remaining outside matchings that form a circuit with $\{i_1, \ldots , i_m\}$.

\subsubsection{Small incompatible sets}
\label{small_incompatible}

Complex paths systems $P_F$ may require many $k$-opt moves to prove $F$ is incompatible with TSP optimality, even when reusing inside matchings.
It is therefore useful to have simple sufficient conditions for incompatibility, allowing us to quickly handle a portion of the Hamilton reveals.

To begin, in one of the first research papers on the TSP, Flood~\cite{flo1956} discusses the case where $F$ consists of two edges. 
\begin{quote}
There is one useful general theorem, which is quickly discovered by each one who considers the traveling-salesman problem.
In the euclidean plane it states simply that the minimal tour does not intersect itself, and this intersection condition generalizes easily for arbitrary [edge lengths] $a_{\alpha\beta}$.
\end{quote}
The general condition is $F = \{ab, xy\}$ is incompatible with TSP optimality if 
\begin{equation}
\label{eqn_2opt}
\max \{d_{ax} + d_{by}, d_{ay} + d_{bx}\} < d_{ab} + d_{xy}.
\end{equation}
This follows from the fact that deleting $ab$ and $xy$ from any tour containing both edges yields two paths that can be reconnected into a tour by adding either $\{ax, by\}$ or $\{ay, bx\}$.
Condition (\ref{eqn_2opt}) implies that, in either case, we improve the length of the tour with the 2-opt move.
We make frequent use of this simple test, checking for incompatible pairs of edges as we consider new Hamilton reveals.

\begin{figure}[htb]
  \centering
  \includegraphics[height=0.8in]{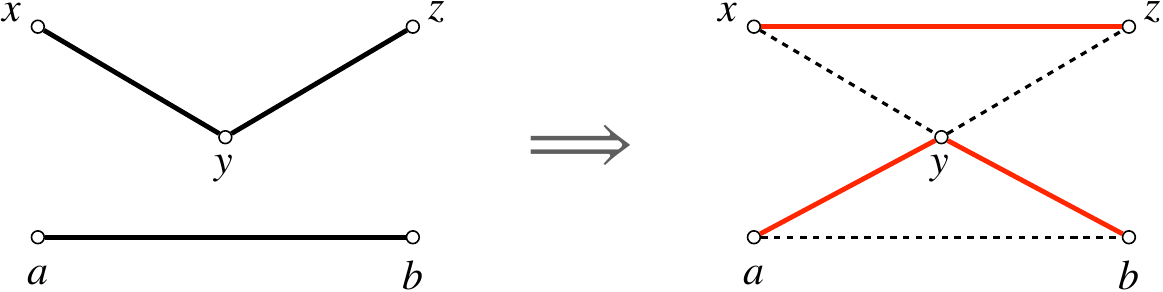}
  \caption{A 3-opt move valid for any tour containing edges $ab$, $xy$, and $yz$.}
  \label{fig_3opt}
\end{figure}

A second simple incompatibility test arises when $F$ consists of three edges $ab$, $xy$, and $yz$. 
Here we assume the five nodes $a, b, x, y, z$ are distinct.
Thus $P_F$ consists of paths $a-b$ and $x-y-z$.
In this case, the condition
\begin{equation}
\label{eqn_3opt}
d_{ay} + d_{by} + d_{xz} < d_{ab} + d_{xy} + d_{yz}
\end{equation}
is sufficient to show $F$ is incompatible with TSP optimality.
Indeed, the single 3-opt move, replacing $F$ by $ay, by,$ and $xz$ is valid for any tour containing $F$.
The move is illustrated in Figure~\ref{fig_3opt}.
Its usefulness comes from the fact that Hamilton reveals, following an isolated-node Tutte move, are paths of length two that can be immediately tested for incompatibility with the edge $ab$ we are trying to eliminate.

\subsubsection{Brute force for small $k$-opt}

For general sets $F$, we search for $k$-opt moves applied to a single tour, rather than aiming for universal conditions such as (\ref{eqn_2opt}) and (\ref{eqn_3opt}). 
In this search, as long as $|F|$ is not too large, small $k$-opt moves can be handled by brute force, considering every subset of $k$ edges to delete from $F$ and examining  all possible ways to reconnect the paths into a tour.

In our implementation, we start with $k = 3$ and proceed to $k = 4$ and $k = 5$ if we do not find a $k$-opt move.
To improve the speed of these computations, at the expense of possibly missing a move, we consider only $k$-opt moves that include deleting the edge $ab$ we are trying to eliminate.

\subsubsection{Adopting a TSP solver}

Moving up to larger values of $k$, we build a TSP instance consisting of nodes
$$V_F = \{v: v \mbox{\ is an end of an edge in}\ F\}$$
and edge lengths obtained from the full TSP instance.
For a specified permutation $\sigma$ and binary vector ${\bf b}$, we create the tour $T_F(\sigma, {\bf b})$ consisting of the paths $P_F$ and the outer matching $O_F(\sigma, {\bf b})$.

Working with this TSP instance, any tour $T'$ that includes all of the edges in $O_F(\sigma, {\bf b})$ and has length less than that of $T_F(\sigma, {\bf b})$ corresponds to a $k$-opt move.
Indeed, we delete the edges $T_F(\sigma, {\bf b})\setminus T'$ and add the edges $T'\setminus T_F(\sigma, {\bf b})$.

Any heuristic or exact TSP code can be used to search for $T'$.
If the code does not directly support the option of fixing edges to be in the tour, a large constant $M$ can be added to the cost of each edge not in $O_F(\sigma, {\bf b})$.
Note that the validity of the elimination procedure is not dependent on the correctness of the TSP solver that is adopted, since the elimination code can easily verify that the $k$-opt move obtained does indeed produce a shorter tour.

In our implementation, we adopt the Held-Karp 1-tree code \cite{hk71} included with the Concorde TSP solver \cite{applegate2006}.
The branch-and-bound search is terminated early if either a tour shorter than $T_F(\sigma, {\bf b})$ is found or a specified maximum number of search nodes is reached.

\subsection{Tutte moves}

Our method for selecting Tutte moves is based on simple heuristic ideas.
First, we prefer nodes $v$ that are in close proximity to the edge $ab$ we are trying to eliminate.
The measure of closeness is either geometric distance to the center of $ab$ or, for non-geometric instances, the minimum number of edges in a path to $a$ or $b$ in the graph $G$.
Secondly, we take a greedy approach, preferring moves that generate a fewer number of Hamilton reveals requiring further Tutte moves to process.
This greedy measure is obtained with a look-ahead routine, selecting a small number of candidate nodes $v$ and counting the number of Hamilton reveals that cannot be discarded using the 2-opt condition (\ref{eqn_2opt}), the 3-opt condition (\ref{eqn_3opt}), and brute-force tests for 3-opt, 4-opt, and 5-opt moves.
Thirdly, to have variety in the search, for each Hamilton reveal we consider both a small number of nodes $v$ that are not contained in the current set $F$ and, if each of these moves returns a failure, a small number of nodes $v$ that are end nodes of paths in $P_F$.

When the initial Tutte move requests only a path of length one, we find it useful to consider a pair of follow-up Tutte moves, selecting as the second and third moves nodes $c$ and $d$, distinct from $a$ and $b$, such that edge $cd$ is either not in the graph $G$ or $cd$ is incompatible with $ab$.
The fact that $cd$ is not available as an edge in an optimal tour adds structure to the corresponding pair of Hamilton reveals and appears to benefit our search process, when compared to making independent choices for $c$ and $d$.
If the small number of pairs of Tutte moves $c$ and $d$ we try each return a failure, then we carry out a second search, requesting an initial Tutte move consisting of an $e$-centered path of length two.

\subsection{Thresholds}

The elimination process requires complexity thresholds on the edge sets $F$ and path systems $P_F$ created by Hamilton reveals.
In our code, this is accomplished by setting $|P_F| \leq 5$ and giving an upper bound on $|F|$ controlled by command-line parameters for the length of the path requested in the initial Tutte move and for the cumulative number of edges added by subsequent Hamilton reveals.

To further restrict the overall complexity of the Hamilton-Tutte trees, for each Tutte move, other than the initial path request, we also bound the maximum number of Hamilton reveals that are not discarded by condition (\ref{eqn_2opt}), condition (\ref{eqn_3opt}), and brute-force tests for 3-opt, 4-opt, and 5-opt moves.
In our code this bound is set to 5, and switched to 1 for tree nodes for which 10 or more edges have been added to $F$, beyond the edges in the initial path.
This practice of switching the bound to 1 permits the code to carry out deep searches in some circumstances.

\subsection{Fixing edges}

\begin{figure}[htb]
  \centering
  \includegraphics[height=1.0in]{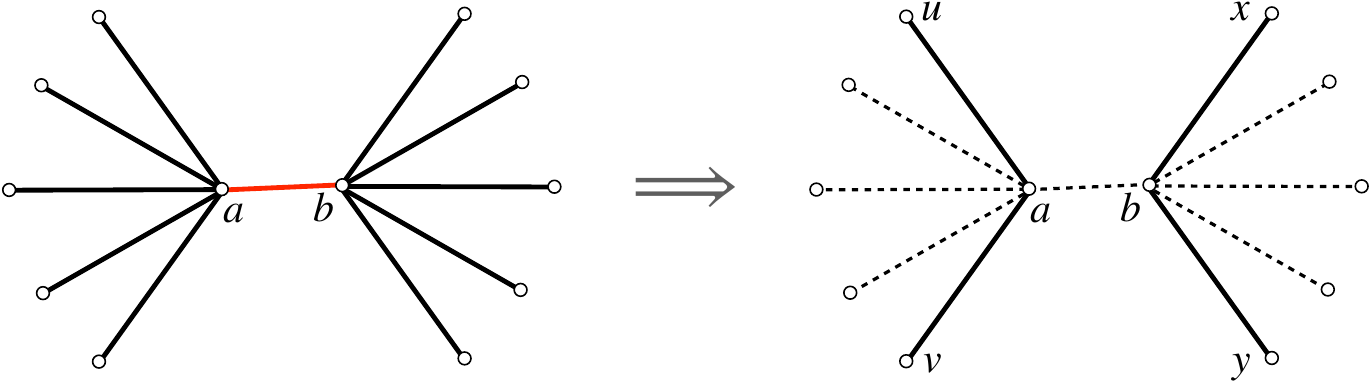}
  \caption{Pair of two-edge paths for fixing edge $ab$.}
  \label{fig_fix}
\end{figure}

The discussion has focused on eliminating edges from the TSP search space, but the techniques can also be applied to fixing edges in the tour, that is, proving an edge $e = ab$ must be present is every optimal tour.
This can be done, for example, by showing that for all pairs of edges $au, av$ meeting node $a$ but not meeting node $b$ and all pairs of edges $bx, by$ meeting node $b$ but not meeting node $a$, the set $\{au, av, bx, by\}$ is incompatible with TSP optimality.
Ruling out all such pairs of two-edge paths through $a$ and $b$, we can conclude that $ab$ must be in every optimal tour.
See the illustration in Figure~\ref{fig_fix}.

\subsection{Non-pairs}

In long computations, it can be beneficial to list pairs of edges $xy, yz$ that are known to be incompatible with TSP optimality.
Hamilton reveals containing such pairs can then be discarded during edge elimination or edge fixing.
To build this {\em non-pair list}, for all nodes $y$ and edges $xy, yz$ meeting $y$, we search for a winning Hamilton-Tutte tree starting with the two-edge path $x-y-z$.
In this setting, the leaf nodes of the game tree correspond to a $\{xy, yz\}$-witness family, defined analogously to the single-edge witness families introduced in Section~\ref{section_witnesses}.

In our implementation of this pair-elimination process, if the initial Tutte move requests only the path of length two, then as the second Tutte move we consider neighbors $v$ of $y$ such that $v \neq x$ and $v \neq z$.
This allows the algorithm to take immediate advantage of the fact that $yv$ cannot be in a tour containing $x-y-z$.

\section{Computation}
\label{section_computation}

The elimination process requires additional choices and settings to produce a working code suitable for large-scale instances. 
In this section we present further details, together with test results using our implementation.

\subsection{Test instances}

Our study focuses on the potential use of edge elimination/fixing to enhance exact solution algorithms for the TSP. 
We thus consider as a test set instances of size beyond the scope of simple optimization methods.
The main source is the standard TSPLIB \cite{rei91, rei95} collection, which contains 13 examples having 3,000 or more nodes, the largest having 85,900.

Each of the large TSPLIB instances is defined by a set of points in the Euclidean plane.
Travel distances are determined either by the {\tt EUC\_2D} norm, defined as the Euclidean distance rounded to the nearest integer, or the {\tt CEIL\_2D} norm, defined as the Euclidean distance rounded up to the next integer.

In addition to the 13 TSPLIB instances, we include in our test set three larger examples, E100k.0, mona-lisa100k, and usa115475, all adopting the {\tt EUC\_2D} norm.
\begin{itemize}
\item E100k.0 contains 100,000 points with integer coordinates drawn uniformly from the 1,000,000 $\times$ 1,000,000 square.
It was created in 2000 by David S. Johnson for the DIMACS TSP Challenge. 
Code for generating the point set is available at \url{http://dimacs.rutgers.edu/archive/Challenges/TSP/download.html}.
\item mona-lisa100k was created in 2009 by Robert Bosch using his TSP Art methods \cite{bosch2019} to render da Vinci's Mona Lisa.
The 100,000-point instance can be downloaded at {\url{https://www.math.uwaterloo.ca/tsp/data/ml/monalisa.html}}.
\item usa115475 consists of the latitude and longitude of 115,475 cities, towns, and villages in the United States.
The point set, created in 2012, is available at {\url{https://www.math.uwaterloo.ca/tsp/data/usa/index.html}}.
\end{itemize}
These three examples have not yet been solved to proven optimality, continuing to provide a challenge for new optimization techniques.

\subsection{Sparse edge sets}

The 16 TSP instances in our test set are defined on complete graphs.
To obtain sparse graphs for the elimination process, we use as a pre-processing routine reduced-cost elimination.
To carry this out, we adopt the LKH and Concorde TSP solvers.
LKH is a TSP heuristic code developed by Helsgaun~\cite{hel2000} and Concorde is the exact solver of Applegate et al.~\cite{applegate2006}, based on the cutting-plane method.

For the 11 instances having fewer than 20,000 nodes, we execute the pre-processing in a ``live'' fashion, simulating the potential use of edge elimination by a TSP solver.
In these cases, we make a run of LKH to produce a high-quality tour and a run of Concorde to produce an LP relaxation.\footnote{LKH version 2.09 was run with settings {\tt CANDIDATE\_SET\_TYPE = POPMUSIC, MAX\_CANDIDATES = 7, RUNS = 1}.
Concorde was run with command-line option -mC48 (to adopt the local cuts procedure up to chunksize 48) and with an option to continue the cutting-plane method until no violated cuts have been added to the LP relaxation.
Due to an excessively long run time, the fl3795 instance was run with Concorde -mC28.}
Reduced-cost elimination is then carried out using the LP relaxation and an upper bound of 1 greater than the length of the LKH tour, to ensure that only edges not in any optimal tour are eliminated.
The results are reported in Table~\ref{tsplib_table}.
\begin{table}[htb]
\caption{LP Reduced-Cost Elimination (LKH+Concorde)}
\label{tsplib_table}
\begin{center}
\begin{tabular}{|c|rrc|rr|}\hline
\multicolumn{1}{|c}{Name} & \multicolumn{1}{c}{LKH} & \multicolumn{1}{c}{Concorde} & \multicolumn{1}{c}{Opt Ratio} & \multicolumn{2}{c|}{LP Edges} \\
\hline
pcb3038 &    163.7s&   5460.2s& 1.00006&   6883&  2.3$n$\\
fl3795  &   4236.2s& 100004.6s& 1.00053&  61609& 16.2$n$\\
fnl4461 &    179.4s&   5074.1s& 1.00004&  10343&  2.3$n$\\
rl5915  &    457.9s&  40867.6s& 1.00011&  29143&  4.9$n$\\
\hline
rl5934  &    490.3s&   9072.4s& 1.00011&  28678&  4.8$n$\\
pla7397 &   9102.7s&  80363.4s& 1.00011&  42180&  5.7$n$\\
rl11849 &   7935.2s&  43422.1s& 1.00010&  78993&  6.7$n$\\
usa13509&   7906.9s&  28402.2s& 1.00009& 136732& 10.1$n$\\
\hline
brd14051&  10510.4s&  61880.5s& 1.00008&  95959&  6.8$n$\\
d15112  &  11117.5s&  38843.6s& 1.00011& 166499& 11.0$n$\\
d18512  &  17449.5s&  56388.8s& 1.00008& 178825&  9.7$n$\\  
\hline
\end{tabular}
\end{center}
\end{table}
The TSPLIB identifier in the  `Name' column includes the number of nodes in the graph.
The running times for LKH and Concorde are reported in seconds on a single core of a linux workstation equipped with two 6-core Intel Xeon CPU E5-2620 2.00GHz processors. 
The `Opt Ratio' column reports the length of the LKH tour divided by the value of the LP relaxation.
We remark that in 7 of the 11 test instances, LKH actually delivered an optimal tour.
Finally, the `LP Edges' column reports the number of edges that were not eliminated via the reduced-cost technique.
For instances over 5,000 nodes, this number ranges from 4.8$n$ up to 11.0$n$, where $n$ denotes the number of nodes in the instance.

For the 5 largest instances in our test set, reduced-cost elimination based on single runs of LKH and Concorde produce very large edge sets that are difficult to manage, both in the algorithm and in the storage of the data.
For these instances, that are either unsolved or tractable only with very long runs of Concorde, we used the best available tour (found by repeated runs of LKH) and the best available LP relaxation (found by runs of Concorde).
The reduced-cost elimination results are reported in Table~\ref{big_table}.
\begin{table}[htb]
\caption{LP Reduced-Cost Elimination (Best Available Tour+LP)}
\label{big_table}
\begin{center}
\begin{tabular}{|c|c|rr|}\hline
%Name & LKH &  Concorde & Opt Ratio & Edges \\
\multicolumn{1}{|c}{Name} & \multicolumn{1}{c}{Opt Ratio} & \multicolumn{2}{c|}{Edges} \\
\hline
pla33810&        1.000164&    1,602,936&   47.4$n$\\
pla85900&        1.000007&      266,273&    3.1$n$\\
E100k.0&         1.000084&    8,280,715&   82.8$n$\\
mona-lisa100k&   1.000019&    1,509,670&   15.1$n$\\
usa115475&       1.000104&   25,009,702&  216.6$n$\\
\hline
\end{tabular}
\end{center}
\end{table}
%The remaining edge sets for these larger instances have up to 216$n$ edges, making a substantial target for the combinatorial edge-elimination process.

\subsection{Computer code}

The elimination process is implemented in 7,781 lines of code written in the C programming language. 
The full code includes also 7,364 lines of code from the Concorde library, handling various utility functions, as well as the Held-Karp 1-tree solver used to check for possible $k$-opt moves.
The code links only the standard C library, allowing it to be run on a wide variety of computing platforms.
The source code is available at {\url{https://github.com/bicobico2/ElimTSP}}.

The main executable function, called {\tt elim}, takes as input a full TSP instance in TSPLIB file format and a sparse graph $G$ expressed as a list of edges.
Optionally, a list of non-pairs, a list of fixed edges, and an input tour can be specified.
These three optional data sets can improve the computation by ruling out certain Hamilton reveals or by reducing the number of edges to process.
In the case of the input tour, the code will not attempt to eliminate any edge that is in the tour and it will not attempt to fix any edge that is not in the tour.

The code is set up to operate in a boss-worker parallel mode.
Workers receive from the boss a small subset of edges $S$ to process, they carry out the elimination algorithm for each edge in $S$, and return to the boss the elimination status for each edge in $S$.
The boss is in a listening mode, responding to worker requests for new subsets of edges to process or to receive the result of worker computations. 
Communication between the boss and workers is enabled by socket-based message-passing routines from the Concorde library.

\subsection{Bootstrapping}

The number of Hamilton reveals in response to a given Tutte move depends directly on the sparsity of the input graph $G$.
This dependency makes it possible to run the elimination process in a bootstrapping fashion, quickly eliminating edges with small Hamilton-Tutte trees in an initial pass, then applying the algorithm again starting with the reduced graph.
This can be done repeatedly, slowly increasing the code's parameters to allow more complex searches for Hamilton-Tutte trees to eliminate additional edges.

In our bootstrapping run, we control the complexity of the Hamilton-Tutte search via the following three settings.
\begin{itemize}
\item[--] {\it neighbors}: the number of nodes, in the proximity of the edge to be eliminated, that are considered as possible candidates for Tutte moves.
\item[--] {\it depth}: the maximum depth of a Tutte node in the Hamilton-Tutte tree.
\item[--] {\it fast}: in fast mode, we test only one candidate $(c,d)$ pair for the first two Tutte moves and we do not employ the TSP solver to search for $k$-opt moves with $k > 5$.
\end{itemize}
In addition, after a sequence of fast elimination rounds, we run a computation to build a non-pair list to improve the speed of further rounds of edge elimination.
Finally, using the reduced edge set and the non-pair list, we run a computation to determine fixed edges, that is, edges that must be in every optimal tour.

The bootstrapping loop proceeds in levels, where the settings are increased to the next level if less than 5\% of the remaining edges are eliminated or less than 25\% of the remaining pairs are eliminated or less than 5\% of the non-fixed edges in the input tour have been fixed.
The level settings we use are displayed in Table~\ref{level_table}.
\begin{table}[htb]
\caption{Settings in Bootstrap Run}
\label{level_table}
\begin{center}
\begin{tabular}{|c|c|c|c|}\hline
%level & Depth &  Neighbors & Witness \\
\multicolumn{1}{|c}{Type} & \multicolumn{1}{c}{Level} & \multicolumn{1}{c}{Depth} & \multicolumn{1}{c|}{Neighbors} \\
\hline
Fast edges  & 1&        2&     5\\
Fast edges  & 2&        2&    10\\
Fast edges  & 3&        3&     5\\
Fast edges  & 4&        3&    10\\
Fast edges  & 5&        4&    10\\
Fast edges  & 6&        4&    15\\ 
Fast edges  & 7&        5&    25\\ 
Fast edges  & 8&        6&    25\\ 
\hline
Non-pairs  & 1&        2&     5\\
Non-pairs  & 2&        3&    10\\
Non-pairs  & 3&        4&    25\\
\hline
Edges & 1&        4&    25\\
Edges & 2&        4&    50\\
Edges & 3&        5&    25\\
Edges & 4&        6&    25\\
Edges & 5&        6&    50\\
\hline
Fixed  & 1&        6&    25\\
\hline
\end{tabular}
\end{center}
\end{table}
A full run adopts eight levels of fast edge elimination, followed by three levels of pair elimination, then five levels of standard edge elimination, and finally one level of edge fixing.

% 48-core network
%   4 compute nodes, two 6-core Intel Xeon CPU E5-2620 2.00GHz processors.
% Boss command line
%   elim -A -h -t TOURS/pcb3038.tour -T DATA/pcb3038.tsp  EDGES/pcb3038.all.edg
% Worker command line
%   elim -g tsp01
% Source code: elim181023.tgz

Computational results for the loop are reported in Table~\ref{loop_full_table}.
\begin{table}[htbp]
\caption{Edge Elimination (48-core network) }
\label{loop_full_table}
\begin{center}
\begin{tabular}{|c|rr|c|rr|rr|r|}\hline
\multicolumn{1}{|c}{Name} & \multicolumn{2}{c}{Fast edges} &  \multicolumn{1}{c}{Non-pairs} & \multicolumn{2}{c}{Edges} & \multicolumn{2}{c}{Fixed} & \multicolumn{1}{c|}{Clock}\\
\hline
pcb3038&  6521&    2.1$n$&  49.4\%&   5548& 1.8$n$&   934& 0.31$n$&    497s \\
fl3795&  21028&    5.4$n$&  20.2\%&  15921& 4.2$n$&   628& 0.17$n$&  27203s \\
fnl4461 &  9928&   2.2$n$&  54.8\%&   9012& 2.0$n$&   973& 0.22$n$&    462s \\
rl5915  & 17582&   3.0$n$&  29.0\%&  11504& 1.9$n$&  3503& 0.59$n$&   7461s \\
\hline
rl5934  & 17114&   2.9$n$&  33.3\%&  12007& 2.0$n$&  3258& 0.55$n$&   6255s \\
pla7397 & 21779&   2.9$n$&  72.6\%&  19989& 2.7$n$&  1484& 0.20$n$&   4339s \\
rl11849 & 38471&   3.2$n$&  35.3\%&  32239& 2.7$n$&  4042& 0.34$n$&   7882s \\
usa13509& 39686&   2.9$n$&  45.1\%&  35211& 2.6$n$&  2950& 0.22$n$&   3844s \\
\hline
brd14051&  43400&  3.1$n$&  50.8\%&  39488& 2.8$n$&  1577& 0.11$n$&   4177s \\
d15112  &  46662&  3.1$n$&  50.7\%&  42271& 2.8$n$&  1528& 0.10$n$&   3893s \\
d18512  &  58537&  3.2$n$&  50.9\%&  58537& 2.9$n$&  1814& 0.10$n$&   5196s \\
pla33810& 143197&  4.2$n$&  38.8\%& 124331& 3.7$n$&  2580& 0.08$n$&  88483s \\
\hline
pla85900&   232768& 2.7$n$& 74.3\%& 214910& 2.5$n$& 17372& 0.20$n$&  29680s \\
E100k.0&    324292& 3.2$n$& 43.9\%& 297603& 3.0$n$& 16879& 0.17$n$&  39065s \\
mona-lisa100k & 299185& 3.0$n$& 80.4\%& 278341& 2.8$n$&   225& 0.00$n$&  13762s \\
usa115475&  374029& 3.2$n$& 45.9\%& 336642& 2.9$n$& 13517& 0.12$n$&  67042s \\
\hline
\end{tabular}
\end{center}
\end{table}
The runs were carried out on a 48-core network of 4 linux compute servers, each equipped with two 6-core Intel Xeon CPU E5-2620 2.00GHz processors.
The `Clock' column reports the wall-clock time in seconds for each test run.
The `Fast edges' and `Edges' columns report the remaining edges after the corresponding rounds of elimination, expressed both as the total number and as a multiple of $n$, the number of nodes in the given instance.
The `Non-pairs' column reports the number of non-pairs, expressed as a percentage of the total number of two-edge paths in the reduced graph.
Finally, the `Fixed' column reports the number of fixed edges, expressed as the total number and as a multiple of $n$.
Averaged over the 16 test instances, the proportion of wall-clock time used in each division was 9\% for fast edges, 5\% for non-pairs, 74\% for edges, and 12\% for fixed.

The elimination runs give only minor reductions for pcb3038, fnl4461, and pla859000, where reduced-cost elimination had already reduced the edge counts to 2.3$n$, 2.3$n$, and 3.1$n$, respectively. 
In all other cases the edge count was reduced by at least a factor of two, resulting in under 3$n$ edges in 11 of these instances, including the three examples having 100,000 or more nodes.
Also, in 14 of the test instances the runs fixed at least 0.1$n$ of the edges.

\subsection{Randomly-generated Euclidean instances}

As a stress-test for the code, we generated 250,000 Euclidean instances, each with $n = 100$ points having integer coordinates drawn uniformly from the 10,000 $\times$ 10,000 square.
The integers were generated as a single stream using the RngStream package written by Pierre L'Ecuyer; see L'Ecuyer \cite{lecuyer1999} and L'Ecuyer et al. \cite{lecuyer2002}.
Starting edge sets for the instances were produced with reduced-cost elimination, using the subtour LP relaxation and upper bound set to 1.001 times the length of a tour found by Concorde's Lin-Kernighan heuristic; these choices create larger edge sets than those produced by Concorde's LP relaxation and an LKH tour, and avoid cases that give a zero integrality gap.

In this test, we did not provide input tours.
This allowed us to check for possible errors in the implementation of the algorithm, running Concorde to find an optimal tour and then verifying that no tour edge was eliminated and only tour edges were fixed.
For elimination and fixing, we again used the bootstrap loop described in Table~\ref{level_table}; this loop is far too time consuming for such small instances, but it allowed us to test the main components of our elimination code.

The results of the test are summarized in Table~\ref{table_r100}.
Each run was carried out on a single hyperthread of an Intel Xeon CPU E5-2620 2.00GHz processor.
Columns `LP Edges', `Reduced', and `Fixed' report the number of edges after reduced-cost elimination, the number edges after the elimination process, and the number of fixed edges, respectively.
The column labeled  `Elim Time' reports the CPU time in seconds for the elimination process.
The rows report mean, median, minimum, and maximum values, taken over the 250,000 instances.
We note that in 1,830 instances, the elimination process was able to fix all 100 edges in the unique optimal tour.
%% tests made elim -A
\begin{table}[htbp]
\caption{250,000 Random Euclidean Instances with $n$ = 100}
\label{table_r100}
\begin{center}
\begin{tabular}{|c|c|c|c|c|c|}\hline
\multicolumn{1}{|c}{Value} & \multicolumn{1}{c}{LP Edges} & \multicolumn{1}{c}{Reduced} & \multicolumn{1}{c}{Fixed} &  \multicolumn{1}{c|}{Elim Time} \\
\hline
Mean &     430.5 &  177.8 &  49.9 &  1014.0s  \\
Median &    403  &  183   &  46   &  1015.7s  \\
Minimum &   159  &  100   &  22   &   186.5s  \\
Maximum &  1673  &  247   & 100   &  3208.5s  \\
\hline
\end{tabular}
\end{center}
\end{table}

In a second test, we generated 250 Euclidean instances with $n = 10,000$ and integer coordinates drawn uniformly from the 10,000,000 $\times$ 10,000,000 square.
For these larger instances, the starting edge set was computed via reduced-cost elimination using a Concorde LP relaxation and an upper bound of 1 greater than the length of an optimal tour.
Again, we did not provide input tours and verified that in each case the elimination results were consistent with an optimal tour computed by Concorde.
The results are reported in Table~\ref{table_r10000}, for runs carried out on a single hyperthread of the Intel Xeon processor.
In each instance, the reduced edge set has under 2.72$n$ edges.
%% tests made elim -A
\begin{table}[htbp]
\caption{250 Random Euclidean Instances with $n$ = 10,000}
\label{table_r10000}
\begin{center}
\begin{tabular}{|c|c|c|c|c|}\hline
\multicolumn{1}{|c}{Value} & \multicolumn{1}{c}{LP Edges} & \multicolumn{1}{c}{Reduced} & \multicolumn{1}{c}{Fixed} &  \multicolumn{1}{c|}{Elim Time}\\
\hline
Mean &     40017.7 &  23305.2 &  2497.7 &  142745.1s \\
Median &   39809   &  23456   &  2438   &  141126.4s \\
Minimum &  22393   &  17482   &  1905   &   52195.2s \\
Maximum &  66318   &  27182   &  4195   &  253582.2s \\
\hline
\end{tabular}
\end{center}
\end{table}

\subsection{Fast elimination}

The bootstrap loop produces strong results, at the expense of considerable computation time.
In potential applications where shorter times are desired, an elimination process can be run with Hamilton-Tutte trees having at most two Tutte moves, taking advantage of the simple path systems that arise in this case.
This is implemented in the {\tt KH-elim} code, available at {\url{https://github.com/bicobico2/ElimTSP}}.

It proved useful to  begin the process with a pass through the full edge set, adopting the Jonker-Volgenant \cite{jv1984} strategy of using 2-opt moves to potentially show very quickly that an edge $(a,b)$ can be eliminated.
In this phase, as a possible single Tutte move, we consider a small number of nodes $y$ in a neighborhood of $(a,b)$.
Rather than considering each pair of edges meeting node $y$ (that is, every Hamilton reply), the Jonker-Volgenant strategy is to make the much quicker check that each individual edge $(x, y)$ meeting $y$ satisfies the incompatibility condition (\ref{eqn_2opt}) with $(a,b)$, thus showing $(a,b)$ can be eliminated.
Following this, we make another pass, considering for each remaining edge $(a,b)$ up to ten candidate $(c,d)$ pairs of Tutte moves, where $c$ and $d$ belong to the neighborhood of $(a,b)$.
In this second phase, we proceed as in the full elimination code, considering all possible Hamilton replies to the $(c,d)$ pair.

Computational results are reported in Table~\ref{single_table} for all test instances having fewer than 20,000 nodes.
%% test made with KH-elim220909 -Jq
\begin{table}[htbp]
\caption{Single $(c,d)$ Pair of Tutte Moves}
\label{single_table}
\begin{center}
\begin{tabular}{|c|rr|rrr|rrr|}\hline
\multicolumn{1}{|c}{Name} & \multicolumn{2}{c}{LP Edges} & \multicolumn{3}{c}{Reduced (1 core)} & \multicolumn{3}{c|}{Reduced (44 cores)}\\
\hline
pcb3038 &    6883&  2.3$n$&   6466& 2.1$n$&  0.0s&  6466& 2.1$n$&  0.1s \\
fl3795  &   61609& 16.2$n$&  31017& 8.2$n$& 50.7s& 31043& 8.2$n$&  2.2s \\
fnl4461 &   10343&  2.3$n$&   9865& 2.2$n$&  0.0s&  9864& 2.2$n$&  0.1s \\
rl5915  &   29143&  4.9$n$&  24596& 4.2$n$&  1.5s& 25107& 4.2$n$&  0.2s \\
\hline
rl5934  &   28678&  4.8$n$&  23426& 3.9$n$&  1.3s& 23448& 4.0$n$&  0.2s \\
pla7397 &   42180&  5.7$n$&  23935& 3.2$n$&  2.9s& 24385& 3.3$n$&  0.4s \\
rl11849 &   78993&  6.7$n$&  61792& 5.2$n$&  6.0s& 61792& 5.2$n$&  0.6s \\
usa13509&  136732& 10.1$n$&  60448& 4.5$n$& 33.5s& 60415& 4.5$n$&  1.4s \\
\hline
brd14051&   95959&  6.8$n$&  60894& 4.3$n$&  9.5s& 60939& 4.3$n$&  0.6s \\
d15112  &  166499& 11.0$n$&  73840& 4.9$n$& 39.6s& 73851& 4.9$n$&  1.9s \\
d18512  &  178825&  9.7$n$&  92220& 5.0$n$& 35.3s& 92279& 5.0$n$&  1.7s \\
\hline
\end{tabular}
\end{center}
\end{table}
The computations were carried out on a linux server equipped with two 22-core Intel Xeon Gold 6238 CPU @ 2.10GHz.
The column labeled ``Reduced (1 core)'' lists results for runs using a single core of the server and the column labeled ``Reduced (44 cores)'' lists results obtained using 44 computation threads, where inter-thread communication is handled by the OpenMP application programming interface. 
For each instance we report the number of remaining edges, the number of remaining edges as a multiple of the number of nodes $n$, and the wall-clock time in seconds.
The run eliminated on average 43\% of the LP edges, in each case in under 3 seconds of wall-clock time when using the full 44 cores of the server.

\section{Certifying results}
\label{section_certification}

A long-term goal of this line of research is to build edge elimination into a tool that can enhance the TSP cutting-plane method, following general ideas outlined in Dantzig et al. \cite{dfj1954, dfj1959}.
The target is to extend the range of current exact solvers, such as the Concorde code.

In this exact-solution setting, it can be helpful to record the Hamilton-Tutte tree that eliminates a specified edge.
The recorded tree can be used in a shorter verification run to certify the edge is not in any optimal tour.
By this means, results can be gathered from remote (possibly non-trusted) computing platforms, using a range of parameter settings or variations of the basic elimination code.
The collected results can then be independently verified, before edges are removed from the cutting-plane computation.

\subsection{Hamilton-Tutte tree storage}

%The key to our verification process is having direct access to the Tutte move needed to counter each Hamilton reveal.
%These moves are conveniently organized in the Hamilton-Tutte tree.

When recording a Hamilton-Tutte tree, we do not store its leaf nodes, since path systems associated with leaf nodes can be shown to be incompatible with TSP optimality without need of additional information.
With each non-leaf tree node $x$, other than the root of the tree, we store the corresponding Tutte move together with the Hamilton reveal associated with the tree edge joining $x$ to $x$'s parent.
Child nodes of $x$ are stored as a list.

Internally, the tree is stored using two types of data structures: {\tt httree} and {\tt htnode}.
An {\tt httree} contains the target type of the tree (edge elimination, edge fixing, or pair elimination) and the algorithm settings and parameters that were used in the tree's construction.
An {\tt httree} contains also a pointer to the {\tt htnode} representing the root of the tree.

An {\tt htnode} for a non-leaf tree node contains the graph node indices for the corresponding Tutte move and, except for the root of the tree, the graph node indices for the edges in the Hamilton reveal that led to the Tutte move.
An {\tt htnode} contains also two (possibly null) pointers, {\tt child} and {\tt next}.
These pointers allow us to store the children of a tree node as a linked list: {\tt child} points to the head of the list, {\tt child}$\rightarrow${\tt next} points to the second node of the list, and so on.

Together with the internal data structures, we need a portable file format for storing and transferring the trees.
For this, the node indices and algorithm parameters are recorded as integers, and are thus easy to store in either text or binary format in a file.
To handle the {\tt htnode} pointers, we assign each node of the tree an integer index and use this integer value in place of pointers to the node.
In the file format, we include as header information the edge or pair we are targeting, together with the algorithm settings.
Then, for each tree node $x$, we have a line containing the index of the parent of $x$, followed by $x$'s index, the Hamilton reveal information, the Tutte move information, and finally the indices of the children of $x$.

%%typedef struct CCelim_htnode {
%%  int hamilton_type;           /* 1:edg 2:3path 3:pair-3path 4:long path */
%%  int hamilton_path_len;       /* length of long path                    */
%%  int hamilton_nodes[CCelim_HTNODE_MAX_PATH_LEN];
%%                               /* node indices for hamilton move         */
%%  int tutte_type;              /* -1:none 0:path 1:end 2:point 3:cd-pair */
%%  int tutte_nodes[2];          /* node indices for tutte move            */
%%  int id;                      /* used when saving elim tree to a file   */
%%  struct CCelim_htnode *child; /* a list of the next hamilton moves      */
%%  struct CCelim_htnode *next;  /* used to link the children              */
%%} CCelim_htnode;

%%typedef struct CCelim_httree {
%%  CCelim_htnode *root;         /* root of the Hamilton-Tutte tree        */
%%  int elimtype;                /* specify edge, fix, or pair             */
%%  int count;                   /* number of nodes in the tree            */
%%  int depth;                   /* depth (max over all nodes) of the tree */
%%  int max_count;               /* upper bound on # nodes in the tree     */
%%  int max_depth;               /* upper bound on depth of the tree       */
%%  int level;                   /* parameter for CCelim_run_elim edge     */
%%  int longpath;                /* parameter for CCelim_run_elim edge     */
%%  int use_tsp;                 /* parameter for CCelim_run_elim edge     */
%%  int max_neighborhood;        /* parameter for CCelim_run_elim edge     */
%%} CCelim_httree;

\subsection{Verification run}

The verification process works through the nodes of the Hamilton-Tutte tree, starting at the root.
The tree nodes are processed in a depth-first order, that is, while processing a node $x$ we recursively process all children of $x$ and then return to continue work at $x$'s parent node.
The process does not depend on the correctness of the computations used to build the tree, but rather uses the stored information to simplify a new elimination run.

To process a tree node $x$, we first list all possible Hamilton reveals in response to $x$'s Tutte move.
This is done as in the full elimination code, using the input graph $G$ and the path system $P_x$ associated with $x$ (that we have by adding $x$'s Hamilton reveal to the path system associated with the parent of $x$).
For each Hamilton reveal $H$ in our list, we first check if there is a child node $c$ with $H$ as its Hamilton reveal.
If there is such a child, we process node $c$ before continuing with the next reveal in our list.
If there is no such child, then we verify that the path system obtained by adding $H$ to $P_x$ is incompatible with TSP optimality.
This last step is done with the techniques used in the main elimination code.

The verification process is simpler and faster than the initial elimination runs, since we avoid the backtracking search for Tutte moves and also avoid tests for incompatibility at non-leaf nodes of the Hamilton-Tutte tree.

\subsection{Sparse edge sets for the 100k-point instances}

We used the verification process in a study of the three largest instances in our test collection, both as a test of the limits of our elimination code and to provide edge sets that can potentially be adopted in exact solution attacks on these currently unsolved examples of the TSP.

The elimination runs made on these instances involved ad hoc changes to the default settings and minor modifications to the source code, exploring alternatives to our standard search directions.
The computations were carried out on a 288-core network of linux compute servers, equipped with two 6-core Intel Xeon CPU E5-2620 2.00GHz processors.
The total computational time (adding together the times on the individual cores) was over 10 core-years in each instance.

The above details make clear this is not a reproducible experiment.
But with the stored Hamilton-Tutte tree for each eliminated edge, for each fixed edge, and for each non-pair, we can verify the final outcome of the many ad hoc runs.

The results are reported in Table~\ref{big_sparse}.
The `Edges' and `Fixed' columns list the number of edges that were not eliminated and the number of fixed edges, respectively.
The `Verification' column reports the running time in seconds to verify the results and the size in gigabytes for the (text) Hamilton-Tutte tree files.
Each of the sparse graphs has average degree under 5.0, improving the average degrees of 6.4, 6.0, and 6.4 in the corresponding results reported in Table~\ref{loop_full_table}.

\begin{table}[htb]
\caption{Sparse edge sets for 100k+ instances}
\label{big_sparse}
\begin{center}
\begin{tabular}{|c|rr|rr|rr|}\hline
\multicolumn{1}{|c}{Name} & \multicolumn{2}{c}{Edges} & \multicolumn{2}{c}{Fixed} & \multicolumn{2}{c|}{Verification} \\
\hline
E100k.0&         241682 & 2.42$n$ & 26106 & 0.26$n$ & 319,731s   & 4.6GB \\
mona-lisa100k&   248972 & 2.49$n$ &  1635 & 0.02$n$ & 189,629s   & 5.5GB \\
usa115475&       286655 & 2.48$n$ & 22394 & 0.19$n$ & 1,950,878s & 6.1GB \\
\hline
\end{tabular}
\end{center}
\end{table}

\section{Remarks}
\label{section_remarks}

We conclude with several observations on the elimination code and its possible application.

\subsection{Code enhancements}

Several ideas, described below, showed promise in our computational tests, but were ultimately removed from the code in an effort to reduce its complexity, particularly in regards to the verification procedure.

\subsubsection{Full-witness families}

In Section~\ref{section_witnesses}, we defined the notion of an $e$-witness family, where $e$ is a target edge to eliminate from a TSP instance specified by a graph $G = (V,E)$ and edge lengths $(d_f: f \in E)$.
We can similarly define a {\em full-witness family} as a family ${\cal Q} = \{F_1, F_2, \ldots, F_q\}$ of edge sets such that, for each $i = 1, \ldots, q$, $F_i \subseteq E$ and every optimal tour $T$ of $G$ has $F_i \subseteq E(T)$ for some $i \in \{1, \ldots, q\}$.
A full-witness family ${\cal Q}$ can be used as a tool for showing multiple edges can be eliminated.
Indeed, to eliminate an edge $f$, it suffices to show that for each $i  = 1, \ldots, q$ the set $F_i \cup \{f\}$ is incompatible with TSP optimality.
Similarly, during the construction of a Hamilton-Tutte tree, an edge set $F$ can be shown to be incompatible with TSP optimality by verifying that for each $i  = 1, \ldots, q$ the set $F_i \cup F$ is incompatible.

To construct a full-witness family, a Hamilton-Tutte tree (with no starting edge or path) can be grown in a breadth-first manner, working to a prescribed upper bound on the number of leaf nodes.
Such trees can be grown in local regions of the graph $G$, for example, by selecting a random set of nodes $S$ and growing a tree for each $s \in S$ with Tutte nodes restricted to a neighbor set of $s$.
The process can be bootstraped, in the sense that previously constructed trees can be used to reduce the size of a new tree, using the incompatibility test described above.

\subsubsection{Order-specific Tutte moves}

Our code for building a Hamilton-Tutte tree follows the outline given in Algorithm~\ref{algo_HT} in Section~\ref{section_nowhere}.
In this algorithm, a Hamilton reveal, producing a path system $P = \{p_1, p_2, \ldots , p_m\}$, leads to a failure if one or more of the $2^{m-1}(m-1)!$ path orderings of $P$ cannot be shown to have a $k$-opt move.
An alternative approach is to make a deeper search for an individual failed path ordering, considering an additional Tutte move to handle only this specific case.

An advantage of using such order-specific Tutte moves is that there are at most $2m$ possibilities for inserting each of the Hamilton reveals into the failed path ordering.
This has the potential for a large savings in computation, if there are few failed path orderings.

\subsubsection{Metric excess}

Hougardy and Schroeder~\cite{hs2014} describe a notion of ``metric excess'' that can be used to show a path of three edges is incompatible with TSP optimality.
The idea is to consider an additional Tutte move $z$, together with a fast mechanism to check that each Hamilton reply to $z$ leads to an improving 3-opt move.
For details see \cite{hs2014}.

\subsubsection{LP reduced costs}

We used LP reduced-cost elimination to create the test sets used in our study, but our elimination code itself does not make use of LP results.
It is natural to consider integrating the two approaches.
For example, during the construction of a Hamilton-Tutte tree, an edge set $F$ is incompatible with TSP optimality if the sum of the reduced costs of the edges in $F$ exceeds $\Delta$, the integrality gap with respect to the LP relaxation and a known TSP tour.
This gives the possibility of avoiding a time-consuming search for a proof that the set is nowhere $k$-optimal.
Or, as a second example, one can target certain edges for elimination or fixing based on reduced costs (large positive values suggest edges for elimination and large negative values suggest edges for fixing).

%%\begin{table}[htb]
%%\caption{Using reduced costs on the 100k+ instances}
%%\label{big_rc}
%%\begin{center}
%%\begin{tabular}{|crc|rc|}\hline
%%\multicolumn{1}{|c}{Name} & \multicolumn{2}{c}{Elim} & \multicolumn{2}{c|}{Elim+RC} \\
%%\hline
%%E100k.0&         +3 edges & 236s &   +3 edges & 241s \\
%%mona-lisa100k&   +5 edges & 122s &  +27 edges & 123s \\
%%usa115475&      +24 edges & 261s & +297 edges & 256s \\
%%\hline
%%\end{tabular}
%%\end{center}
%%\end{table}

%%
%%
\subsection{Using elimination in TSP algorithms}

We describe below several possibilities for exploiting the elimination routine.

\subsubsection{LP relaxations over sparse edge sets}

Unlike reduced-cost elimination, our combinatorial routines can remove edges that take on positive values in an optimal solution to an LP relaxation, potentially improving both the LP bound and the computational performance of an LP solver.
To illustrate this, in Table~\ref{improve_lp} we present LP information for the three 100k+ instances.
In each of the examples, the ``(full)'' values refer to optimizing over the complete graph and the ``(sparse)'' values refer to optimizing over the sparse edge sets with fixed edges set to one, using the results described in Table~\ref{big_sparse}.
\begin{table}[htb]
\caption{Improved LP optimality ratios/gaps with sparse edge sets}
\label{improve_lp}
\begin{center}
\begin{tabular}{|c|c|c|c|}\hline
LP Relaxation & E100k.0 & mona-lisa100k & usa115475 \\
\hline
Subtour (full)         &    1.0064790 & 1.0009764 & 1.0073171 \\
Subtour (sparse)       &    1.0058662 & 1.0009670 & 1.0066773 \\
Gap Closed             &        9.4\% &     0.9\% &     8.7\% \\ \hline
Concorde (full)        &    1.0000848 & 1.0000193 & 1.0001036 \\
Concorde (sparse)      &    1.0000826 & 1.0000193 & 1.0001022 \\
Gap Closed             &        1.6\% &     0.0\% &     1.4\% \\ \hline
Concorde+Cuts (full)   &    1.0000837 & 1.0000193 & 1.0001028 \\
Concorde+Cuts (sparse) &    1.0000796 & 1.0000193 & 1.0000951 \\
Gap Closed             &        4.9\% &     0.0\% &     8.2\% \\ \hline
Current Best           &    1.0000486 & 1.0000127 & 1.0000803 \\
\hline
\end{tabular}
\end{center}
\end{table}

The first block of rows in the table reports results for the standard {\em subtour relaxation} of the TSP.
This LP is the starting point for TSP cutting-plane methods, going back to the work of Dantzig, Fulkerson, and Johnson \cite{dfj1954}.
%%Let $x$ denote the vector of variables $(x_e: e \in E)$, for any set $S \subset V$ let $\delta(S)$ denote the set of edges having exactly one end in $S$, and for any $F \subseteq E$ let $x(F)$ denote $\sum(x_e: e \in F)$.
%%With this notation, the subtour relaxation can be written as follows.
%%\begin{eqnarray*}
%%& \min \sum(d_ex_e: e \in E)\\
%%& x(\delta(\{v\})) = 2, \ \ \forall \; v \in V\\
%%& x(\delta(S)) \geq 2,  \ \ \forall \; \emptyset \neq S \subsetneq V\\
%%& 0 \leq x_e \leq 1,    \ \ \forall \; e \in E.
%%\end{eqnarray*}
The first two rows list the optimality ratio provided by the relaxation, that is, the value of the best known tour divided by the value of the dual LP solution.
The ``Gap Closed'' row reports the percentage of the integrality gap closed using the sparse edge sets.
The second block of rows reports the corresponding results for the best LP relaxations produced by the Concorde code working with the full edge sets.
These relaxations were obtained using multiple runs of Concorde, following the strategies outlined in Section 16.4 of Applegate et al.~\cite{applegate2006}.
The dual LP values were again improved for E100k.0 and usa115475, but the sparse edge set did not improve the bound for the Mona Lisa instance.

In the third block of rows, we applied Concorde's cutting-plane routines to the LP relaxations reported in the second block.
In these tests, we used the command-line parameter ``-mC100'' to repeatedly run Concorde's cutting-plane loop, in each iteration increasing by 4 the value of $t_{max}$, which controls the size of chunks considered in Concorde's local-cuts routine.
(See Section 16.1 of Applegate et al.~\cite{applegate2006}.)
The runs for E100k.0 and mona-lisa100k finished in under 70 hours, while the two runs for usa115475 were terminated after 320 hours (with the full instance reaching $t_{max} = 44$ and the spare instance reaching $t_{max} = 40$).

Finally, the ``Current Best'' row reports the optimality ratio for the current best LP relaxations, found using the reduced edge sets and multiple additional runs of Concorde.
We must note that, whereas the first three blocks demonstrate improvements due only to edge elimination, the better bounds reported in this final row are due also in part to new cutting-plane search techniques that were applied to the sparse LP relaxations.
Indeed, it is difficult to separate the contribution of the sparse graphs.
Nonetheless, the large improvement over the best full-graph results are suggestive of the potential for application of the elimination routine.

An important observation is that, when comparing the full-graph runs and the sparse-graph runs from the third block of rows in Table~\ref{improve_lp}, there was a significant reduction in the percentage of CPU time spent in solving the LP relaxations arising in the cutting-plane process, as reported in Table~\ref{percent_lp}.
\begin{table}[htb]
\caption{Percentage of computation time used by the LP solver}
\label{percent_lp}
\begin{center}
\begin{tabular}{|c|c|c|c|}\hline
LP Relaxation & E100k.0 & mona-lisa100k & usa115475 \\
\hline
Concorde+Cuts (full)   &    29.4\% & 69.9\% & 97.5\% \\
Concorde+Cuts (sparse) &    25.1\% & 57.1\% & 93.0\% \\
\hline
\end{tabular}
\end{center}
\end{table}
This would contribute to the improved current bounds obtained in computations carried out with the reduced edge sets.

\subsubsection{General MIP techniques}

With a sufficiently sparse edge set, it is possible to apply general MIP techniques to improve an LP relaxation for a TSP instance.
This makes it possible to combine the combinatorial cutting planes studied in the TSP literature and the algebraic methods that have been very successfully applied in MIP solvers.
Note, however, that for exact TSP solutions it is important to adopt numerically safe MIP cutting planes and bounds, such as those described in \cite{cdfg2009} and \cite{eifler2023}.

\subsubsection{Planar graphs}

Another direction is to use the elimination process to produce graphs having structural properties that can be exploited in TSP solution methods.
For example, the reduced graph for mona-lisa100k is planar, opening up several possibilities for improving the execution of the cutting-plane method.

First, for planar graphs, Letchford~\cite{le00} described a polynomial-time separation algorithm for an important class of TSP inequalities called domino-parity constraints.
The implementation of his algorithm by Cook et al.~\cite{ceg2007} required a heuristic method for extracting a planar graph from an LP solution, but this step can now be avoided in the Mona Lisa example, resulting in exact separation for this class of constraints.

Second, the solution set of the subtour relaxation of the TSP is a polytope in $\mathbb{R}^{|E|}$, called the {\em subtour polytope}, defined by exponentially many linear constraints.
For planar graphs, Rivin~\cite{rivin2003} gave an extended formulation of the subtour polytope, showing it is a projection of a polytope in $\mathbb{R}^{3|E|}$ having no more than $|V| + 6|E|$ constraints;
a nice description of Rivin's formulation can be found in Pashkovich~\cite{pash2012}.
For Mona Lisa, we created an integer-programming model consisting of the Riven system together with the constraints from the current best LP relaxation, reported in Table~\ref{improve_lp}.
Using the IBM CPLEX 22.1.0.0 mixed-integer programming solver, a branch-and-bound run (after 7,342 search nodes) reported a lower bound of 5,757,132, potentially closing 19.2\% of the integrality gap if the computation could be carried out with numerically safe methods.
%Although this is not an actual lower bound for mona-lisa100k, the CPLEX run indicates a possible use of Rivin's formulation.

\subsubsection{Edge sets in local-search heuristics}

A reduced edge set can also be adopted in heuristic-search methods, serving as a candidate set of edges for inclusion in a TSP tour.
Indeed, the best-known tour for the E100k.0 instance was found by Keld Helsgaun with LKH in 2013, using results of an early version of our elimination code.
This computation reduced the tour length from 225,786,958 to 225,784,127.
The reduction of 2,830 units improved the gap between the best-known tour length and current LP bound by over 20.5\%.
Note that in this context it is also possible to employ non-exact methods for reducing an edge set, such as in the study by Fischer and Merz~\cite{fischer2007}.

\end{document}